\documentstyle[preprint,aps]{revtex}

\begin{document}
\long\def\cut#1{}
\begin{center}

\Large

{\bf  Straightening of Thermal Fluctuations in Semi-Flexible
 Polymers by Applied Tension}

\large \vspace{0.5cm}
{\sl Udo Seifert$^1$,  Wolfgang Wintz$^1$, and  Philip Nelson$^2$}
\vspace{0.5cm}
\normalsize

$^1$Max-Planck-Institut f\"ur Kolloid- und
Grenz\-fl\"achenforschung, Kantstrasse 55, 14513 Teltow-Seehof, Germany;
%
%\large\vspace{0.5cm}{\sl Philip Nelson} \vspace{0.5cm}
%\normalsize
%
$^2$Physics and Astronomy,
University of Pennsylvania, Philadelphia, PA 19104 USA
%\vspace {1cm}
\end{center}
%DRAFT, \ \ \today.
%?82.70-y, % Disperse systems
%?87.22.Bt. % Membrane and subcellular physics and structure
\begin {abstract}
We investigate the propagation of a suddenly applied tension along a thermally
excited semi-flexible polymer using analytical approximations,
scaling arguments
and numerical simulation. This problem is inherently non-linear. We find
sub-diffusive propagation with a dynamical exponent of 1/4. By generalizing
the internal elasticity, we show that tense strings exhibit qualitatively
different tension profiles and propagation  with an exponent of 1/2.

\end{abstract}

%\eject

\def\beq{\begin{equation}}
\def\ee{\end{equation}}
\def\pcite{\protect\cite}
\def\p{\partial}
\def\g{\gamma}
\def\Tk{T/\kappa}
\def\kT{\kappa/T}
\def\eps{\epsilon}
\def\u{u_{lm}}
\def\c{{\rm const}}
\def\d{\dot}
\def\dd{\ddot}

\def\lsim
{\protect \raisebox{-0.75ex}[-1.5ex]{$\;\stackrel{<}{\sim}\;$}}

\def\gsim
{\protect \raisebox{-0.75ex}[-1.5ex]{$\;\stackrel{>}{\sim}\;$}}

\def\lsimeq
{\protect \raisebox{-0.75ex}[-1.5ex]{$\;\stackrel{<}{\simeq}\;$}}

\def\gsimeq
{\protect \raisebox{-0.75ex}[-1.5ex]{$\;\stackrel{>}{\simeq}\;$}}

\def\r{{\bf r}}
\def\f{{\bf f}}

\def\pcite{\protect \cite}

Characteristic for soft matter systems such as polymers or membranes is the
often subtle interplay between energy and entropy \cite{dege92,safr94}.
Thermal motion determines the conformations of these systems
crucially. For instance, the typical end-to-end distance even of a
semi-flexible
polymer (let alone a Gaussian polymer) is much smaller than its contour
 length. Likewise for vesicles, thermal
fluctuations can store a significant part of the ``true'' surface area.
{}From the perspective of
the projected length or area, this part is hidden. For vesicles or membranes,
the hidden area can be pulled out without stretching the true area by
application of a localized force such as the suction pressure in
a micropipet \cite{evan90} or the action of optical tweezers \cite{z:barz95a}.
For polymers, recent advances in experimental techniques
using magnetic beads, optical tweezers or flow fields
made feasible detailed experimental studies of the conformations of single long
molecules
\cite{schu90,smit92,perk95,wirt95,bens95,cluz96}). With these techniques it has
become possible to measure the force-extension curve for these soft
systems and to compare them with theoretical predictions
\cite{broc93,broc95,mark95,kroy96}. These
studies focus on the equilibrium conformations under the action of
a {\em stationary} force.

The purpose of this letter is to investigate the {\em dynamics} of the
straightening of initial thermal fluctuations. The approach to
the stationary state  takes
time because  these objects are
immersed in a viscous medium (usually water).
Since we are interested in the principle mechanism, we will simplify
the system as much as possible
still retaining what we believe to be the salient features. Rather than
studying
the complicated  dynamics of a two-dimensional membrane in three dimensional
space, we will, therefore,
study the straightening for a one-dimensional semiflexible chain confined to
a two dimensional world. The unwinding of a Gaussian polymer in
a homogeneous flow is  the only other system so far for which the dynamical
approach to the new equilibrium has been studied using scaling arguments
\cite{broc95}.
In contrast to this work, the conformation of our chains are
dominated by bending modes.
Also, we will
consider the case of pulling one end holding the other end fixed,
unlike the free-end situation in a homogeneous flow  considered in
\cite{broc93,broc95}. Our problem is thus analogous to
the case of pulling on a vesicle containing a fixed
volume of fluid \cite{z:barz95a}.

Our main observations are that tension spreads slowly in a long
chain, with a scaling law depending on the elastic character. We give
numerical results along with simple scaling arguments to explain them.

{\sl Model:\ \ } To make the calculation tractable we will neglect the
thermal fluctuations in our dynamics: their only role is to prepare a
rough initial state at time $t=0$, in which the total contour length
per unit of projected
real-space length is greater than one.  We also
assume that at $t>0$ the bending stiffness $\kappa$ is negligible compared to
the applied force $\f$; the only role of $\kappa$ is then to set the initial
roughness of the contour (see below).

Let us write the shape of a ``polymer'' confined to 2d as
$\r(s)=(x(s),y(s))$ along its arc length $0\leq s\leq L$. We fix the right end
to $\r(L)={\bf 0}$. At the left end, a force $\f=-f{\hat{\bf x}}$ pulls in
the  $-x$ direction.

The total energy is then given by
\beq
F\equiv -\f\ \r(L)+{1\over 2} \int_0^L ds \gamma(s)\dot\r(s)^2
\ee
where dots denote derivatives with respect to arc length.
The Lagrange multiplier $\gamma(s)$, which can be interpreted as the
local tension, is necessary to preserve the
arc length constraint $\dot \r(s)^2=1$ which arises from the
inextensibility of the polymer.
We assume local isotropic dissipation and  ignore
long range hydrodynamic interactions.
 The equation of motion is then
\beq
\Gamma^{-1}\p_t\r=-\delta F/\delta \r=\p_s(\gamma\dot \r ) +
(\f+\gamma\dot \r )\delta(s)
\label{eq:mo}
\ee
where $\Gamma$ is an inverse friction coefficient.

There is no explicit equation of motion for the Lagrange multiplier
 $\gamma$.
This quantity has  to be determined from
\beq(1/2\Gamma)\p_t\dot \r^2=\d\r\p_s^2(\g\d\r)=
 \ddot \g-\g\ddot \r^2=0
\label{eq:gamma}.
\ee

At any time $t$, this relation for $\g$ is a second order differential equation
which requires for its solution two boundary conditions. The first, at
$s=0$, is
\beq
\gamma(0)=f
\label{bc0}
\ee
The fixed boundary condition at $s=L$ requires $(1/\Gamma)\p_t\r(L)
=\d\g(L)\d\r(L)+\g(L)\dd\r(L)=0$. This relation implies both
\beq
\d\gamma(L)=0
\label{bc5}
\ee
 and $\dd\r(L)=0$ because $\d\r \dd\r=0$.
The tension profile $\g(s)$ can thus be calculated
for any instantaneous conformation. This profile can then be put into the
equation of
motion for $\r$ which can then be integrated one time step.

{\sl Linear theory:\ \ }
Even for our simplified model, the coupled non-linear equations
(\ref{eq:mo}) and (\ref{eq:gamma}) cannot be  solved
analytically. Let us consider first the linearized
problem, which will fail in an instructive way.  Choose a single mode
initial conformation of the form $y(s,t=0)  = a_q^0 \sin qs $.
Note that $x(s)$ follows once we specify $y(s)$ because of the
arclength constraint.

Ignoring the non-linear term,
the equation for the tension becomes $\dd \g =0$ which leads
with the boundary conditions to the flat profile
$\g(s) = f$. Thus the tension spreads instantaneously in this
approximation. In the linearized  equation of motion for $y$,
the modes decouple and one obtains  the usual relaxation form,
$a_q (t)=a_q^0e^{-\Gamma f q^2 t}$.

The failure of the linear theory can be seen  clearly by looking at the
first correction term.
Writing $\gamma(s)=f + \g_1(s)$, we have
 $\dd\gamma_1(s)=f \dd \r^2$, which implies
for the single mode configuration
 $\gamma_1(L)\sim -  L^2 f {|a_q^0|}^2 q^4$.  This is less than the
lowest order term $f$ only if
\beq
L\ll1/ q^2|a_q^0|  .
\label{eq:L}
\ee Thus depending on the
wave-length and the amplitude of the initial conformation, there
is a maximal chain length over which the linearized
theory can be applied. For any longer chain, the problem becomes
inherently non-linear.

{\sl Simulations:\ \ }
To gain insight into the non-linear problem, we simulated a discretized
version of the equations of motion.
The initial conformation is written as
\beq
y(s,0) =\sum_q a_q^0 \sin qs
\ee
with $q=n\pi/L, n\geq 1$.
This choice guarantees that $y(0,0)=y(L,0)=0$. For simplicity, we also
restrict $y(0,t)=0$ for all times so that the left endpoint moves
only horizontally \cite{foot2}.

The small initial amplitudes  are chosen from a Gaussian distribution
with width
\beq\langle {|a_q^0|}^2\rangle=T/\kappa L q^4 ,
\label{eq:init}
\ee
where $T$ is temperature in units of energy and $\kappa$ is the
bending stiffness.

In Fig.1, we show ten snapshots
of the configuration and the corresponding tension profiles.
The tension spreads inward, decreasing the amplitudes
in this range. For a quantitative analysis, we define a penetration length
$\xi(t)$ over which the tension has already spread. For concreteness
we will define $\xi$ as the point where
$\gamma(\xi)=f/2$. The initial value and the
time evolution of this penetration length depends on the initial
conformation. Averaging 50  runs, we find the power law
\beq
\langle\xi(t)\rangle \sim t^z
\ee
where the dynamic exponent $z=0.24\pm0.01$. We will now show how this
empirical scaling law follows from a simple physical picture.

{\sl Scaling argument I:\ \ }
An exact solution of the non-linear
equations of motions looks impossible. Good insight, however,
can be gained from the structure of the equation for the tension
profile (\ref{eq:gamma}). Since $\dd \g$ has the same sign as $\g$, the
two boundary conditions (\ref{bc0}) and (\ref{bc5})
 imply that $\g(s)$ is
positive, monotonically decreasing and convex \cite{foot3}.
 We will now seek an
approximate effective  equation for the time development of $\g$.
For uniform tension the modes decay as
$a_q(t)=\exp(-\Gamma \gamma q^2 t) a_q^0  $. Since this says that the
short-wavelength modes decay the fastest, we will make the
``adiabatic'' Ansatz that
$\g$ varies slowly in space compared to the wavelengths $2\pi/q$ of
the relevant modes; thus we write
\beq
y(s,t)=\sum_q a_q(s,t) \sin qs
\ee and  regard
$a_q$ as slowly-varying functions of arclength $s$ and of time:
\beq
a_q(s,t)=\exp(-\Gamma \gamma(s,t) q^2 t) a_q^0.\ee
Using eq.~(\ref{eq:init}), the average time-dependent
local curvature then
can be approximated as \cite{foot4}
\beq
\langle\dd y(s,t)^2\rangle = \sum_q {T\over \kappa L q^4}q^4
\exp(-2\Gamma \g(s,t) q^2 t)
{}.
\ee
Written in this form, the curvature $<\dd y(s,t)^2>$
depends only implicitly on the arc
length $s$ via the function $\g(s,t)$. With this simplification,
 Eq. (\ref{eq:gamma}) for the unknown time dependent tension acquires the
simple form
\beq
\dd \gamma (s) = \gamma  \sum_q {T\over \kappa L q^4}q^4 \exp(-2\Gamma \g(s)
q^2 t)
\equiv
-{\p V(\g)\over \p \g}.
\ee

The equation for the tension profile
has thus become  a simple mechanical equation for the one-dimensional motion
of a particle with unit
mass in a potential $V(\g)$  where $\gamma$ plays the role of position
and the original spatial variable $s$ plays the role of time. The
original time $t$ becomes a mere parameter. The potential exhibits two
regimes,
\beq
V(\gamma) \approx
 \left\{\matrix{
\displaystyle - \ \g^2 , & \g\ll \g_c \cr
\noalign{\medskip}
\displaystyle - \ {T\over \kappa  (\Gamma t)^{1/2}}\gamma^{3/2} , &
 \g \gg  \g_c \cr
 }\right .
\ee
 separated by a cross-over tension $\g_c\equiv 1/\Gamma q_m^2 t$.
Here, $q_m$ is a high momentum cut-off for the $q$-modes.

The particle starts  with  ``initial'' condition $\gamma(0)=f$
 and has to reach
 zero velocity
at ``time'' $s=L$ because of
$\d \g(L)=0$. In the limit $L\to \infty$, it then follows that the
 particle has zero total energy
from which we obtain the initial condition
$\d\g(0)=  -2(V(f))^{1/2}<0$. Identifying
$\d\g(0)$ with $-f/\xi(t)$, we obtain
\beq
\xi(t) \sim (\kappa/T)^{1/2}(\Gamma ft)^{1/4}
\ee in good agreement with our numerical simulation.

For a rough estimate of the relevant scales, let us ignore all factors of
order unity. For the friction coefficient $\Gamma$ we can take the
inverse of the bulk viscosity $\eta$, so
 $\Gamma=1/\eta =100 $cm$^3$/erg sec.
A typical weak force is $f=0.1$ pN \cite{smit92}.
For DNA, $\kappa/T\simeq 100$nm \cite{smit92} and thus
$\xi(t)\simeq 1\mu$m$(t/$sec$)^{1/4}$.
For actin,  $\kappa/T\simeq 10\mu$m \cite{actin},
so $\xi(t)\simeq 10\mu$m$(t/$sec$)^{1/4}$.
Thus, the dynamics of straightening in biopolymers
 should be accessible
to video microscopy techniques.

{\sl Generalization:\ \ }
\def\A{{\cal A}}
In the analysis above we took the initial chain configuration to be
governed by bending modes. Another case of interest is when a floppy chain is
initially under tension, and the tension is suddenly increased at time
zero. To cover both cases let us replace
the sample for the initial amplitudes  in Eq. (\ref{eq:init}) by
%\beq
$\langle {{|{a_q^0}|}}^2\rangle=\A^{3-2b}/Lq^{2b} $,
%\label{eq:init2}
%\ee
where $\A$ has the dimension of a length. Then the case $b=2$, where
$\A=\kappa/T$, was discussed above, while
$b=1$ (where $\A=T/\Sigma$) corresponds to a string with
tension $\Sigma$.

Our scaling argument can be repeated for general $b>1/2$. One finds the
effective potential
\beq
V(\g)\sim-\A^{3-2b}( \Gamma f t)^{(2b-5)/2}
\gamma^{b-1/2}
 \ee for $ \g\gg  \g_c $.
This potential leads to a penetration length
%\beq
$\xi(t)\simeq t^{(5-2b)/4} $.
%\ee

We have studied the case $b=1$ numerically. Here we find a dynamical exponent
$z=0.50\pm 0.01$ clearly different from  the prediction $z=3/4$
which would follow from the above scaling argument.

What went wrong?
Closer inspection of the tension profiles indicates a qualitative
difference between the tense string $b=1$ and the semi-flexible $b=2$ cases.
For the string, the 10 snapshots shown in Fig. 2 and the
corresponding tension profiles reveal that the
tension profile is not  exponential but rather decays quite linearly
over a long region, and the moving boundary of this region remains
sharply defined. Intuitively the difference stems from the fact that
more of the initial excess length is in short-wavelength modes in the
tense string case. Since these modes are damped the fastest, the
string straightens immediately when it feels the tension, and so
most of the resistance to straightening comes from pulling a straight
string through a viscous medium, which gives a linear tension
profile. The straight region ends at a point controlled by the total
length pulled so far. In the semirigid
case, most of the excess length is initially in long-wavelength modes,
which decay slowly. The tension propagates forward before the string
has a chance to straighten, and so its front is not so well-defined.

{\sl Scaling argument II:\ \ }We can turn these words into another simple
scaling argument.
 We separate the string into two parts. In the
left, {straightened,} part of length $\xi(t)$, we assume that
all fluctuations have
already been pulled out. Then the tension profile becomes
%\beq
$\g(s)=f-\d\g(L) s $.
%\ee
In the right, {unperturbed,} part, pulling has not yet had a
significant effect. Thus the tension profile for $s>\xi(t)$ is exponential,
%\beq
$\g(s)=\g(\xi(t))\exp(-(s-\xi(t))/\xi_{init})$,
%\ee
with a {\it
time-independent} penetration length
$\xi_{init}\equiv ({1/\dd \r^2})^{1/2}$ as given by the initial
configuration. Matching the two profiles so that  $\d \g$ is
continuous at $s=\xi(t)$, we obtain
\beq
\gamma(\xi(t))=f/
(1+\xi(t)/\xi_{init})\simeq f\xi_{init}/\xi(t) .
\ee
The rate $d\xi(t)/dt$ with which the straightened regime grows is
proportional to the force $\gamma(\xi(t))$ with which this
regime ``pulls'' at the left end of the unperturbed regime.
We thus obtain % the relation
%\beq
$d\xi(t)/dt \sim \gamma(\xi(t))\sim f\xi_{init}/\xi(t) $.
%\ee
This equation is readily solved to give
\beq
\xi(t) \sim (ft)^{1/2} .
\ee This ``diffusive'' behavior indeed corresponds to our simulation
results for $b=1$. We can easily understand it in terms of the
intuitive picture sketched above: the total friction on the straight
segment is proportional to its length $\xi(t)$, and so the string
velocity is $v\sim \xi^{-1}$. Letting $\alpha$ be the
initial excess contour length divided by $L$, conservation of string
says that the front velocity is
$\d\xi=v/\alpha\sim\xi^{-1}$. Solving this equation then reproduces
$\xi\sim\sqrt t$.

In Fig.3, we show the numerically determined exponents for various values
of $b$.  We have strong numerical evidence that $z=1/2$
for $b\leq 1$ and $z<1/2$ for $b>1$.
This plot shows that the two scaling theories yield upper limits
to this exponent.

{\sl Conclusion:\ \ }
We have seen how thermal fluctuations, together with a viscous
surrounding medium,
impede the transmission of a suddenly imposed tension in a polymer. As we
pointed out, this phenomenon is inherently nonlinear. In a model
neglecting stochastic noise, simulations show that the tension
propagation in a stiff chain is subdiffusive with an exponent
$1/4$. We showed how this law also follows from
a simple self-consistent scaling theory. We also generalized both the
model and the scaling theory to the case of
chains with arbitrary initial
elasticity. For tense strings, the spreading becomes
diffusive, in the sense that the tension is nonnegligible in a region
of size $\sim\sqrt t$.
The application of these ideas to the case of suddenly-pulled
membranes \cite{z:barz95a} will be addressed in future work.

\medbreak{\noindent\sl Acknowledgments:\ \ }
We are grateful to H.-G. D\"obereiner and J. Krug for helpful discussions.
This work was supported in part by the
US/Israeli Binational Foundation grant 94--00190 and NSF grant
DMR95--07366.

\begin{figure}[b]
\caption{(top) 10 subsequent snapshots of conformations $(x(s),y(s))$ of
a semi-flexible chain $(b=2)$ with total
chain length 250. The snapshots are taken at equidistant times.
The inset shows how
 two adjacent bends are straightened.
(bottom) Corresponding tension profiles as a function of the arc length.}
\end{figure}

\begin{figure}[b]
\caption{Same as Fig.1 for a string ($b$=1).}
\end{figure}

\begin{figure}[b]
\caption{Dynamical exponent $z$ as a function of $b$. Error bars
have the size of the symbols. The lines refer to the two scaling
estimates.}
\end{figure}


\begin{thebibliography}{10}
\bibitem{foot2}
Strictly speaking, this choice implies that the  boundary condition
$\g(0)=f/\d x(0)$ replaces (\protect\ref{bc0}).
\bibitem{foot3}
\mbox{For a constant curvature, the tension profile would
be}
$\g(s)=f\cdot{\cosh((L-s)/\xi)/ \cosh(L/\xi)}$,
with a penetration length $
\xi =\sqrt {{\dd \r^2}}$.
\bibitem{foot4}
{For the simple scaling argument to come,
 we can safely replace $\dd \r^2=\dd y^2
/(1-\d y^2)$ by $\dd y^2$.}
\end{thebibliography}
\end{document}